\documentclass[ps,prd,preprint,superscriptaddress,preprintnumbers,eqsecnum,showpacs,nofootinbib,nopacs,nobibnotes]{revtex4}
\usepackage{amsfonts,bm}
\usepackage{graphicx}
\usepackage{pstricks}
\usepackage{color}

\newcommand{\be}{\begin{equation}}
\newcommand{\bea}{\begin{eqnarray}}
\newcommand{\ee}{\end{equation}}
\newcommand{\eea}{\end{eqnarray}}

\def\s#1{{\scriptscriptstyle #1}}

\def\1eq#1{Eq.~(\ref{#1})}
\def\2eqs#1#2{Eqs.~(\ref{#1}) and~(\ref{#2})}
\def\3eqs#1#2#3{Eqs.~(\ref{#1}),~(\ref{#2}) and~(\ref{#3})}
\def\noeq#1{(\ref{#1})}

\def\fig#1{Fig.~\ref{#1}}

\def\tr{\mathrm{Tr}\,}

\def\G{\Gamma}
\def\cw{c_\s{W}}
\def\sw{s_\s{W}}
\def\MW{M_\s{W}}
\def\MZ{M_\s{Z}}
\def\MX0{M_\s{X_0}}

\def\n#1{({\it #1}\,)}

\def\ie{{\it i.e.},~}
\def\eg{{\it e.g},~}
\def\vev{{\it vev}}

\begin{document}

\title{Scalar Resonances in the Non-linearly Realized\\
Electroweak Theory}
\date{October, 2012}

\author{D. Binosi}
\email{binosi@ect.it}
\affiliation{European Centre for Theoretical Studies in Nuclear
  Physics and Related Areas (ECT*) and Fondazione Bruno Kessler, \\ Villa Tambosi, Strada delle
  Tabarelle 286, 
 I-38123 Villazzano (TN)  Italy}
 
\author{A. Quadri}
\email{andrea.quadri@mi.infn.it}
\affiliation{Dip. di Fisica, Universit\`a degli Studi di Milano via Celoria 16, I-20133 Milano, Italy\\
and INFN, Sezione di Milano, via Celoria 16, I-20133 Milano, Italy}

\begin{abstract}
\noindent
We introduce a physical scalar sector in a SU(2)$\otimes$U(1) electroweak theory in which the gauge group is realized non linearly. By invoking  
theoretical as well as experimental constraints, we build a phenomenologically viable model in which a minimum of four scalar resonances  appear, and the mass of the CP even scalar is controlled by a vacuum expectation value; however, the masses of all other particles (both matter as well as vector boson fields) are unrelated to spontaneous symmetry breaking and generated by the St\"uckelberg mechanism. We evaluate in this model the CP-even scalar decay rate to two photons
and use this amplitude to perform
 a preliminary comparison with the recent LHC measurements.  
As a result, we find that the model exhibits a preference for a negative Yukawa coupling between the top quark and the CP-even resonance.
\end{abstract}


\maketitle

\section{Introduction}

The experimental programme for probing the Spontaneous Symmetry Breaking (SSB) mechanism of the Standard Model (SM) of particle physics has recently witnessed a major breakthrough with the simultaneous announcements  by the ATLAS~\cite{:2012gk} and CMS~\cite{:2012gu} collaborations of the observation of a new bosonic particle with a mass
of about $125$~GeV.

Though one cannot claim yet any significative discrepancy between the properties of this new particle and the expectations for the SM Higgs field, there is a $2\sigma$ evidence for an enhanced $\gamma\gamma$ channel (and  slightly suppressed $WW$ and $ZZ$ channels), which, in a SM-like scenario, can be accounted for by a modified (possibly negative) Yukawa coupling and a moderate rescaling of the Higgs to vectors coupling~\cite{Giardino:2012dp,Ellis:2012hz,Espinosa:2012im}. To be sure,  the LHC measurements of these processes will significantly 
improve in the near future, thus leading either to a full confirmation
of the SM scenario or to the discovery of new physics beyond it. However, given the present situation, it is particularly important to compare the experimental data against all possible theoretically sound scenarios that can account for possible deviations of the particle couplings from the SM results.

A relatively unexplored model in this context is an electroweak theory in which the SU(2)$\otimes$U(1) gauge group is realized non-linearly. In fact, the usual Higgs mechanism~\cite{Englert:1964et,Higgs:1964ia,Higgs:1964pj,Guralnik:1964eu} is  based on a linear representation 
of the gauge group: Masses are generated by SSB through the 
appearance of a non-zero vacuum expectation value (\vev) of a
physical scalar field, triggered by the usual quartic (mexican hat) potential.
On the other hand, in a model in which the gauge group is realized non-linearly, masses are generated via the St\"uckelberg mechanism~\cite{Stueck}, that is 
through the coupling with the flat connection of the gauge group.
As a consequence, the couplings of a scalar resonance would not be related to the masses of the particles  which it couples to,
unlike those of the Higgs field(s) in the SM and 
extensions thereof.

In this paper we discuss in detail  
how one can include scalar resonances in the nonlinearly realized electroweak theory
within the formalism based on the Local Functional Equation (LFE)~\cite{Bettinelli:2008ey,Bettinelli:2008qn,Bettinelli:2007tq,Bettinelli:2009wu,Ferrari:2005va,Ferrari:2004pd,Bettinelli:2007kc,Ferrari:2005ii,Quadri:2010uk}.
We will analyze what properties emerge for these particles in such a scenario. We will refer to this model as the Non Linear Standard Model (NLSM).

Clearly, the nonlinearity of the gauge transformation implies that
the model is not power-counting renormalizable; however, 
the severe ultraviolet (UV) divergences of the Goldstone fields are maintained under control by means of the LFE, which encode in a mathematically rigorous way the non-trivial deformation of the 
(non-linearly realized) gauge symmetry, induced by radiative corrections.

In addition, perturbation theory can  be still organized in the number of loops 
by exploiting the so-called Weak Power-Counting (WPC) condition~\cite{Ferrari:2005va,Bettinelli:2008qn,Bettinelli:2007tq}.

 The WPC requires that
only a finite number of ancestor amplitudes (\ie amplitudes without external
Goldstone legs) exists at each order in the loop expansion
 and therefore it is the strongest requirement one can ask for
once (strict) power-counting renormalizability is relaxed~\cite{Quadri:2010uk}.
 It is therefore a reasonable criterion for building a model
in the presence of a nonlinearly realized gauge theory, where power-counting
renormalizability does not hold.
 Moreover, in the formulation based on the LFE
the divergences of amplitudes with at least one Goldstone leg are uniquely fixed by the LFE itself.

 A distinctive feature of nonlinearly realized gauge
theories based on the WPC is the appearance of two independent mass
parameters~\cite{Bettinelli:2008qn,Quadri:2010uk} for the $W$ and $Z$ bosons.
This holds true also for a grand-unified SU(5) nonlinearly realized gauge
model, as recently shown in~\cite{Bettinelli:2012jv}. 

 It turns out that the WPC requirement  provides  strong constraints on the possible
terms in the tree-level action and on the matter content of the theory,
when scalar resonances are introduced.

The main results of the paper are the following
\begin{itemize}

\item Unlike in effective electroweak theories~\cite{Giudice:2007fh,Contino:2010mh,Grober:2010yv}, no scalar singlet is allowed
in the NLSM, the minimal choice of physical scalar fields being an SU(2) doublet,
corresponding to four particles: two neutral (one CP-even, $\chi_0$, and one CP-odd, $\chi_3$)
and two charged physical $\chi^\pm$ resonances.

\item SSB, triggered by a suitable quartic potential, must occur for the SU(2) doublet along the $\chi_0$-component, \ie $\chi_0 = v + X_0$; the reason  is that otherwise one cannot accommodate for the suppression of the decay width of  $X_0 \rightarrow \gamma \gamma$
with respect to (w.r.t.) the decay modes $X_0 \rightarrow VV$, $V=W,\ Z$ (which, without SSB, would be radiatively generated as well).

\item However, the \vev~ of the scalar field has nothing to do with the masses of the other physical particles
in the model\footnote{It rather controls the strength of the tree-level generated partial widths $X_0 \rightarrow VV$, $X_0 \rightarrow VV^*$ and $X_0 \rightarrow V^*V^*$.}, since the latter are induced via the St\"uckelberg mechanism.
This property makes the comparison with the linear theory particularly
interesting, since one can work in a SSB scenario where the couplings 
are not directly related to the masses of the particles. 

\item From the phenomenological point of view, in the NLSM there are non-standard one-loop contributions to the UV finite and gauge invariant partial width $\G({X_0 \rightarrow \gamma \gamma})$ coming from charged scalar resonances; however, we find that these cannot explain by itself the non-standard best fits to LHC results, and  that a negative Yukawa coupling is also needed.

\end{itemize}

The paper is organized as follows. In Section~\ref{NLSMbuilding} we introduce the non-linearly realized electroweak theory and we sketch how scalar resonances can be introduced preserving WPC. We also construct the mass term for all relevant spin 0, spin $1/2$ and spin 1 particles as well as the couplings. 

A fully detailed discussion of the formal properties of the theory
(including its BRST quantization and the appropriate gauge condition respecting the LFE) will be given elsewhere; here we rather focus on the results
that this analysis would lead to.

Next in Section~\ref{pheno} we discuss the phenomenological signatures of the model, and in particular show, through the analysis of the $X_0$  amplitude  to two photons, that 
 a negative value of the 
 $X_0\overline{t}t$ Yukawa coupling is preferred. The paper ends with some conclusions (Section~\ref{concl}), and an Appendix showing, for the readers convenience, the so-called bleached version for all the relevant  physical fields
of the theory.

\section{\label{NLSMbuilding}Building up the Non Linear Standard Model}

In the nonlinearly realized electroweak theory of \cite{Bettinelli:2008ey,Bettinelli:2008qn} the usual SM gauge bosons and fermions are supplemented with an SU(2) matrix $\Omega$ which contains the Goldstone fields~$\phi_a$, reading
\bea
\Omega = \frac{1}{f} (\phi_0 + i g \phi_a \tau_a); \qquad
\phi_0 = \sqrt{f^2 - g^2 \phi_a^2} \, .
\label{a.1}
\eea
The trace component $\phi_0$ is a solution of the nonlinear
SU(2) constraint
\bea
\phi_0^2 + g^2 \phi_a^2 = f^2 ,
\label{a.2}
\eea
where $f$ is a parameter with the dimension of a mass, that, being unphysical, must cancel in any physical NLSM amplitude. Under $\rm SU(2)_\s{L} \otimes U(1)_\s{Y}$ the matrix $\Omega$ transforms as
\bea
\Omega' = U \Omega V^\dagger; \qquad
U \in {\rm SU(2)_\s{L}}\, ,  V = \exp \left(i \frac{\alpha}{2} \tau_3\right) \in {\rm U_\s{Y}(1)}, 
\label{t.1}
\eea
where $\tau_i$ ($i=1,2,3$) are the usual Pauli matrices.

From the original gauge bosons and fermion fields, one can construct the so-called bleached $\rm SU(2)_\s{L}$ variables~\cite{Bettinelli:2008qn,
Bettinelli:2007tq}, \ie $\rm SU(2)_\s{L}$ gauge-invariant combinations in one-to-one correspondence with the original fields. For the reader's convenience we collect the bleached counterparts of the gauge boson and fermion fields in Appendix~\ref{app:a}. Notice that, due to $\rm SU(2)_\s{L}$ invariance, the hypercharge
of the bleached variables equals their electric charge, in agreement with the Gell-Mann-Nishijima formula.

The model is clearly not power-counting renormalizable; however for an appropriate choice of the tree-level interaction vertices the 
WPC holds~\cite{Ferrari:2005va,Bettinelli:2008qn},
and only a finite number of divergent 1-PI ancestor amplitudes exists at every order in the loop expansion.
On the other hand, already at the one-loop level there is an infinite
number of divergent 1-PI Goldstone amplitudes~\cite{Ferrari:2005va,Bettinelli:2007tq,Bettinelli:2008qn}; they are however uniquely constrained by the 1-PI ancestor amplitudes
through the LFE~\cite{Ferrari:2005ii} which controls the deformation of the classical non-linearly realized gauge symmetry
induced by radiative corrections~\cite{Bettinelli:2007kc}. 

Also it should be stressed that the theory fulfills physical unitarity
(\ie cancellation of intermediate unphysical states in the physical
amplitudes), as a consequence of the validity of the Slavnov-Taylor
identity~\cite{Ferrari:2004pd}.

Since there are of course many possible nonlinear realizations of the
electroweak theory  (for instance electroweak effective theories,
based on a low-energy expansion~\cite{Buchmuller:1985jz}), we would like to make some more comments
on the WPC, which will be used as a model-building principle in what follows.

The WPC condition can be given a clean mathematical interpretation
as a criterion for choosing uniquely the (decorated) 
Hopf algebra of the model~\cite{Connes:1999yr,Connes:2000fe}. This is because the graphs in the expansion 
based on the {\em topological} loop number are uniquely identified by the WPC
 itself~\cite{Ferrari:2005va}. 

Indeed, while the number of divergences increases order by order in the loop expansion, the results in~\cite{Bettinelli:2008ey,Bettinelli:2008qn,Bettinelli:2007tq,Bettinelli:2009wu,Ferrari:2005va,Ferrari:2004pd,Bettinelli:2007kc,Ferrari:2005ii,Quadri:2010uk} guarantee however that the theory can be still made finite by a finite number of counterterms, order by order in the loop expansion. This means that there exists a suitable exponential map~\cite{EbrahimiFard:2010yy} on the given Hopf algebra of the theory allowing the removal of all the divergences.

On the other hand, the addition of finite higher orders counterterms entails a change of the Hopf algebra of the theory. 
Therefore, this condition singles out the nonlinearly realized electroweak theory based on the WPC from other possible nonlinear realizations: while the nonlinearly realized theory controlled by the WPC is a genuine loop expansion (based on a given Hopf algebra of the model, defined by the WPC itself), effective field theories in the low energy expansion are not, having a different Hopf algebra.

The Lagrangian ${\cal L}_\s{\rm NL}$ of the nonlinearly realized electroweak
theory is highly constrained by WPC~\cite{Bettinelli:2008qn}.
The latter requires the
self-couplings between gauge bosons 
as well as the couplings between gauge bosons and fermions
be the same as the SM ones~\cite{Bettinelli:2008ey,Bettinelli:2008qn}.
However, the tree-level Weinberg relation does not hold
in the nonlinear theory\footnote{In this respect it should be noticed that for a linearly realized electroweak group the Weinberg relation still holds if one only imposes WPC (as opposed to strict power-counting renormalizability)~\cite{Quadri:2010uk}.}, and an
independent mass parameter $\kappa$ arises 
in the vector boson sector; this fact yields a different relation
between the mass of the $Z$ and $W$~\cite{Bettinelli:2008ey}, and namely
\be
\MZ^2 = (1 + \kappa) \frac{\MW^2}{\cw^2}.
\ee
In the above equation $\cw$ is the cosine of the
Weinberg angle $\theta_W$; the latter is defined
according to the usual relation 
\be
\tan \theta_\s{W} = \frac{g'}{g},
\ee
where $g$ and $g'$ are respectively the $\rm SU(2)_\s{L}$ and $U(1)_\s{Y}$ coupling
constants.
The existence of the second mass parameter $\kappa$ is
a peculiar feature of the nonlinearly realized electroweak theory~\cite{Bettinelli:2009wu};  notice that $\kappa$ is related to the usual $\rho$ parameter~\cite{Ross:1975fq} through
\be
\frac1\rho=1+\kappa.
\ee

\subsection{No SU(2) scalar singlet allowed}

We can now extend the field content of the nonlinearly realized electroweak theory by adding physical scalar fields. 

The simplest possibility would be to consider an additional neutral, CP-even SU(2)-singlet field $h$.
This choice is commonly made in the nonlinear low-energy effective Lagrangian parameterizing the electroweak symmetry breaking sector
\cite{Giudice:2007fh,Contino:2010mh,Grober:2010yv}
\bea
{\cal L}_\s{\rm eff} & = & \frac{1}{2} \partial_\mu h \partial^\mu h - V(h)
+\frac{v^2}{4} \left ( w^+ w^- + \frac{1}{2}(1+\kappa) w_3^2 \right )
\left [ 1 + 2 a \frac{h}{v} + b \frac{h^2}{v^2} +
b_3 \frac{h^3}{v^3} + \dots \right ] \nonumber \\
& & - \frac{v}{\sqrt{2}} 
\left [ \widetilde{\overline u}^i_\s{L} y^u_{ij} u^j_\s{R} 
+ \widetilde{\overline d}^i_\s{L} y^d_{ij} d^j_\s{R} \right ]
\left ( 1 + c \frac{h}{v} + c_2 \frac{h^2}{v^2} +\dots \right ) + {\rm h.c.},
\label{no.1}
\eea
where all the symbol appearing are described in Appendix~\ref{app:a}.
Notice that, if the custodial symmetry is {\it imposed} 
(\ie, $\kappa=0)$, the gauge boson mass term in \1eq{no.1} reduces to 
\bea
\left . \frac{v^2}{4} \left[w^+ w^- + \frac{1}{2}(1+\kappa) w_3^2 \right]
\right |_{\kappa=0} = 
\frac{v^2}{4} \tr (D_\mu \Omega)^\dagger D^\mu \Omega,
\label{no.2}
\eea
which represents the familiar form used, \eg in~\cite{Ellis:2012hz,Espinosa:2012im}.
In eq.(\ref{no.2}) $D_\mu$ is the covariant derivative w.r.t.
the $\rm SU(2)_\s{L} \otimes U(1)_\s{Y}$ gauge group:
\bea
D_\mu \Omega = \partial_\mu \Omega - i g A_\mu \Omega
- i g' \Omega \frac{\tau_3}{2} B_\mu.
\label{s.5}
\eea

\begin{figure}[!t]
\includegraphics[scale=0.55]{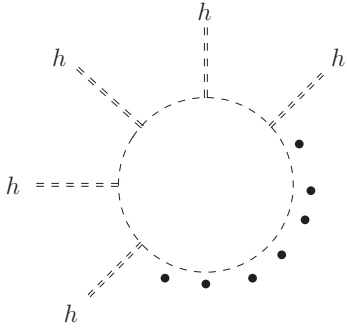}
\caption{\label{WPC}Ancestor amplitude with external $h$ legs, generated by the interaction term $w^+w^-h$, leading to the maximal violation of WPC due to the presence of the $h$-Goldstone interaction vertex $h\partial^\mu\phi\partial_\mu\phi$}.
\end{figure}

However the interactions between the gauge bosons and the scalar $h$
in the first line of \1eq{no.1} are forbidden by WPC, since they
give rise to vertices $V \sim h \partial_\mu \phi \partial^\mu \phi$
with one $h$ and two Goldstone legs.
Already at the one-loop level, these vertices gives rise to divergent 1-PI amplitudes
with an arbitrary number of external $h$-insertions (see~\fig{WPC}),
leading to a maximal violation of  WPC.

\subsection{SU(2) scalar doublet}

The next option is to consider an SU(2) doublet of physical scalars:
\bea
\chi = \frac{1}{\sqrt{2}} ( \chi_0 + i \chi_a \tau_a ).
\label{s.1}
\eea
In order to determine the $\chi$-dependence of the
classical action allowed by WPC, we first consider the sector spanned 
by the kinetic terms and the scalar-gauge bosons interactions, and list below all possible CP-even and neutral 
gauge-invariant operators of dimension $\leq 4$
that can be obtained from the bleached variables of \1eq{a.4}.  The kinetic terms are 
\be
\begin{array}{cccc}
   \partial_\mu \widetilde{\chi}_0 \partial^\mu \widetilde{\chi}_0;\qquad &
   \partial_\mu \widetilde{\chi}_3 \partial^\mu \widetilde{\chi}_3; \qquad& 
   {\cal D}_\mu \widetilde{\chi}^+ {\cal D}^\mu \widetilde{\chi}^-,
\end{array}
\label{s.2.1}
\ee
where ${\cal D}_\mu$ denotes the photon covariant derivative. 
The trilinear couplings involving a gauge field are
\be
\begin{array}{cccc}
   i w_{3\mu} \widetilde \chi^+ {\cal D}^\mu \widetilde \chi^- + {\rm h.c.};\quad&
    i w_{3\mu} \widetilde \chi_0 \partial^\mu \widetilde \chi_3 + {\rm h.c.};\quad&
    %i w_{3\mu} \widetilde \chi_3 \partial^\mu \widetilde \chi_0\\
    i w^+_\mu \widetilde \chi_0 {\cal D}^\mu \widetilde \chi^- + {\rm h.c.};\quad &
    i w^+_\mu \widetilde \chi_3 {\cal D}^\mu \widetilde \chi^- + {\rm h.c.},
\end{array}
\label{s.2.2}
\ee
while those involving two vector bosons and a scalar are
\be
\begin{array}{cccc}
   \widetilde{\chi}_0 w_3^2;\qquad &
   \widetilde{\chi}_0 w^+ w^-.
\label{s.2.3}
\end{array}
\ee
Finally the quadrilinear couplings are given by
\be
\begin{array}{cccccc}
   \widetilde{\chi}_0^2 w_3^2;\qquad& 
   \widetilde{\chi}_0^2 w^+ w^-;\qquad&
   \widetilde{\chi}_3^2 w_3^2;\qquad&
   \widetilde{\chi}_3^2 w^+ w^-;\qquad&
   \widetilde{\chi}^+\widetilde{\chi}^- w_3^2; \qquad&
    \widetilde{\chi}^+\widetilde{\chi}^- w^+ w^-.
\label{s.2.4}
\end{array}
\ee

Enforcing WPC requires to take a linear combination of these monomials
such that all the interaction vertices with two
Goldstone fields, two derivatives and any number
of other (non-Goldstone) legs vanish. It turns out that there is just one combination of this kind  (with canonically normalized $\chi_a,\chi_0$ fields), and namely
\bea
\frac{1}{2} %\int\!\diff^4x \,  
\tr (D^\mu \chi)^\dagger (D_\mu \chi) \, .
\label{s.4}
\eea

Notice in particular that the trilinear scalar-vector-vector
couplings of \1eq{s.2.3} are not allowed.

This fact has clearly some important phenomenological consequences. Specifically, the decays $\chi_0 \rightarrow ZZ$, $\chi_0 \rightarrow WW$ are radiative processes, and therefore one cannot account for their experimentally measured enhancement w.r.t. the diphoton decay channel $\chi_0 \rightarrow \gamma \gamma$. Thus one is forced to introduce SSB in the scalar resonance sector. 

This is achieved by adding the (usual) gauge-invariant quartic potential 
\be
V(\chi)  =  - \frac{\lambda}{16} \left[ \tr (\widetilde \chi^\dagger \widetilde \chi)^2\right]^2 + \frac{ \mu^2 }{2} ~\tr (\widetilde \chi^\dagger \widetilde \chi)
= - \frac{\lambda}{16} \left[\tr (\chi^\dagger  \chi)^2\right]^2 + 
\frac{ \mu^2}{2} \tr (\chi^\dagger  \chi).
\label{qpot.1}
\ee 
After the $\chi_0$ component has acquired a \vev~, $\chi_0 = v + X_0$, the linear term in $V(\chi)$ disappears when the condition $\mu^2 = \lambda v^2/4$ is satisfied, while at the same time a mass term for $X_0$ arises, with $\MX0 = \sqrt{2} \mu$. However notice that in the model at hand, $v$ is completely  unrelated to the masses of the other particles (fermions and  gauge bosons), rather controlling the (tree-level) strength
of the decay rates of $X_0$ in two $W$'s and two $Z$'s. 

Let us end this section by noticing that the UV degree of the fields $X_0,\chi_3,\chi^\pm$ is not changed by the introduction of the
potential~\noeq{qpot.1}; then, this allows us to add to the Lagrangian
two independent mass terms for $\chi_3$ and $\chi^\pm$ through their bleached counterparts
\bea
\frac{1}{2} M^2_3 \widetilde \chi_3^2 + M^2_\pm \widetilde \chi^+ \widetilde \chi^-,
\label{qpot.2}
\eea
without altering the unit UV degree of the  $\chi$ doublet fields.

\subsection{Gauge bosons mass terms}

As a consequence of SSB, induced by the potential in \1eq{qpot.1},
the $W$ and $Z$ bosons acquire masses as in the SM.
However, in the NLSM two independent mass invariants can be added.
They implement the mass generation through the St\"uckelberg mechanism
and can be written concisely as follows \cite{Bettinelli:2008ey}.
We define
\bea
\Omega_{\alpha \beta} = \frac{1}{f} \Phi^\s{C}_\alpha \Phi_\beta;
\qquad \Phi = \pmatrix{ i \phi_1 + \phi_2 \cr \phi_0 - i \phi_3};\qquad\Phi^\s{C} = i \tau_2 \Phi^* = \pmatrix{ \phi_0 + i \phi_3 \cr i \phi_1 - \phi_2}.
\label{gbm.1}
\eea
Then the following independent gauge-invariant combinations can be added to the classical action without violating the WPC:
\bea
\frac{1}{2} \frac{A}{f^2}(D_\mu \Phi)^\dagger (D^\mu \Phi) + \frac{1}{4} \frac{B}{f^2} (\Phi^\dagger D_\mu \Phi)^\dagger (\Phi^\dagger D_\mu \Phi).
\label{gbm.2}
\eea
The first term gives mass to both the $W$ and the $Z$ while respecting the custodial symmetry, the second one only to the $Z$. The coefficient $B$ 
measures the strength of the violation of the tree-level Weinberg relation
and is thus expected to be small.

If one chooses
\bea
A =\frac{4 \MW^2}{ g^2} - v^2; \qquad
Bf^2 = \frac{8 \MZ^2}{g^2 + {g'}^2} - 
\frac{8 \MW^2}{g^2},
\label{gbm.3}
\eea
the two independent parameters $\MW$ and $\MZ$ can be directly 
identified with the tree-level masses of the 
$W$ and the $Z$ vector bosons (the SM limit corresponding clearly in this case to the condition $A \rightarrow 0$ and $B \rightarrow 0$).
We remark that, as LHC data accumulate, one expects to be able to
probe the validity of custodial symmetry within a suitably chosen benchmark parameterization for the fit to the LHC experimental results~\cite{LHCHiggsCrossSectionWorkingGroup:2012nn,Passarino:2012cb}, thus obtaining direct information on the $B$ parameter introduced above.

The terms contributing to the masses of the gauge bosons are
\bea
{\cal L}_\s{\rm m} =% \int\! \diff^4x \, 
%\left [   
\frac{1}{2}  \tr (D_\mu \chi)^\dagger (D^\mu \chi) +
\frac{1}{2} \frac{A}{f^2} (D_\mu \Phi)^\dagger (D^\mu \Phi) + \frac{1}{4} \frac{B}{f^2} (\Phi^\dagger D_\mu \Phi)^\dagger (\Phi^\dagger D_\mu \Phi) 
%\right]
.
\label{mass.1}
\eea
It is convenient to rescale the Goldstone fields with $\phi^\pm \rightarrow \frac{\sqrt{A}}{f} \phi^\pm,
\phi_3 \rightarrow \frac{\sqrt{C}}{f} \phi_3$, \mbox{$C = A + \frac{B}{2}f^2$},
in order to get canonically normalized $\phi$'s.
Then the Goldstone bosons and the fields describing the physical resonances
are obtained by means of an orthogonal transformation, mixing
the fields $\phi^\pm, \phi_3$ and $\chi^\pm, \chi_3$ as follows
\bea
{\phi^\pm}' = \frac{1}{\sqrt{1 + \frac{A}{v^2}}} \left (
\chi^\pm + \frac{\sqrt{A}}{v}  \phi^{\pm} \right ); &\qquad&
%\qquad
{\chi^\pm}' = \frac{1}{\sqrt{1+\frac{A}{v^2}}} \left (
- \frac{\sqrt{A}}{v}  \chi^\pm + \phi^\pm \right ); \nonumber \\
\phi_3' = \frac{1}{\sqrt{1+\frac{C}{v^2}}} \left ( \chi_3 
+ \frac{\sqrt{C}}{v} \phi_3 \right ); &\qquad&
%\qquad
X_3' = \frac{1}{\sqrt{1+\frac{C}{v^2}}} \left ( -\frac{\sqrt{C}}{v} \chi_3 + \phi_3 \right ).
\eea
The $\chi'$ are invariant under the linearized gauge transformations, as it should be for physical scalars, while the $\phi'$ are not, being
the (unphysical) Goldstone bosons of the theory~\footnote{A detailed
treatment of the BRST quantization of the model and of the appropriate choice of the gauge-fixing condition in order to preserve the LFE will be presented elsewhere.}.
Notice that in the limit $A \rightarrow 0$ and $B \rightarrow 0$ the number of
degrees of freedom changes: the $\chi$ become the Goldstone fields,
as is evident from \1eq{mass.1}, while all beyond-the-SM resonances
disappear from the ${\cal L}_\s{\rm m}$. 

\subsection{Yukawa couplings and Flavour Changing Neutral Currents suppression}

The most general parameterization of the interaction
of the physical scalars $\chi$  with two fermions
%, compatible with WPC, 
is 
\bea
{\cal L}_\s{\rm{S \overline f f }} & = & \widetilde{\chi}_0 \left ( \widetilde{\overline u}^i_\s{L} y^u_{ij} u^j_\s{R}  +
\widetilde{\overline d}^i_\s{L} y^d_{ij} d^j_\s{R} \right ) +  \widetilde{\chi}_3 \left ( i \widetilde{\overline u}^i_\s{L} y^{'u}_{ij}  u^j_\s{R}  +
i \widetilde{\overline d}^i_\s{L} y^{'d}_{ij}   d^j_\s{R}  + {\rm h.c.} \right ) 
+ \widetilde{\overline u}^i_\s{L} y^{ud}_{ij}  d^j_\s{R} \widetilde \chi^+ + {\rm h.c.},\nonumber \\
\label{s.6}
\eea 
with fermion masses  generated by the bleached combinations presented in \1eq{a.g.7} of Appendix~\ref{app:a}. 
%With the choice of eq.(\ref{}) the fields $\widetilde u_\s{L}, \widetilde d_\s{L}, u_\s{R}$ 
%and $d_\s{R}$ are  mass eigenstates. 

Within this general choice, a finite number of divergent 1-PI ancestor amplitudes arises order by order in the loop expansion for arbitrary matrices $y^u, y^d, y^{'u}, y^{'d}$ (with an UV index $1/2$ for both the fermions and the fields $\chi$). 
This not very satisfactory,  since in this case flavour changing neutral
currents (FCNCs), mediated by a neutral scalar boson, are not in general suppressed. 

A natural mechanism for forbidding FCNCs  in
the nonlinear theory is based on an extended symmetry
for the composite operators appearing in~\1eq{s.6}. 
In order to formulate it, let us introduce the 
external sources $Y^u_{ij}, Y^d_{ij}, Y^{'u}_{ij}, Y^{'d}_{ij}$
with couplings
\bea
 \widetilde{\chi}_0 \left ( \widetilde{\overline u}^i_\s{L} Y^u_{ij} u^j_\s{R}  +
\widetilde{\overline d}^i_\s{L} Y^d_{ij} d^j_\s{R} \right ) 
+  \widetilde{\chi}_3 \left ( i \widetilde{\overline u}^i_\s{L} Y^{'u}_{ij}  u^j_\s{R}  +
i \widetilde{\overline d}^i_\s{L} Y^{'d}_{ij}   d^j_\s{R}  + {\rm h.c.} \right ) 
+ \widetilde{\overline u}^i_\s{L} Y^{ud}_{ij}  d^j_\s{R} \widetilde \chi^+ + {\rm h.c.}
\eea
By imposing that all the interaction vertices involving the Goldstone fields $\phi$'s,
one source $Y$ and two fermion legs vanish, we single out the unique  combination
\bea
{\cal L}_\s{\rm S \overline f f } & = & 
 (\overline{Q}_\s{L})^i Y^{u}_{ij} u^j_\s{R} \Xi^C +
 (\overline{Q}_\s{L})^i Y^{d}_{ij} d^j_\s{R} \Xi + {\rm h.c.},
\label{fcnc.2}
\eea
where we have introduced the left SU(2) doublet
\be
(Q_\s{L})^i = \pmatrix{u_\s{L}^i \cr d_\s{L}^i},
\label{fcnc.1}
\ee
and set
\be
\Xi = \pmatrix{ i \chi_1 + \chi_2 \cr \chi_0 - i \chi_3}; \qquad
\Xi^C = i \tau_2 \Xi^* =  \pmatrix{ \chi_0 + i \chi_3 \cr i \chi_1 - \chi_2},
\label{theXis}
\ee
from which one can write the SU(2) doublet of ~\1eq{s.1} in the compact form
\be
\chi_{\alpha \beta} = \Xi^C_\alpha \Xi_\beta.
\ee
Notice that the emerging structure~\noeq{fcnc.2} 
implements the suppression of scalar boson mediated FCNCs as in the 
SM, through the extension to the scalar sector of the celebrated GIM mechanism~\cite{Glashow:1970gm}.

The sources $Y$ acquire UV degree $1$, which is the maximum value
they can get, since at one-loop there are fermion loops with two
external $Y$ sources leading to Feynman amplitudes with 
superficial degree of divergence $2$. 

Trilinear scalar-fermion-fermion couplings are next assumed to be
generated by a shift $Y^{u}_{ij} \rightarrow Y^{u}_{ij} + y^{u}_{ij}$,
$Y^{d}_{ij} \rightarrow Y^{d}_{ij} + y^{d}_{ij}$.
Then the interaction~\noeq{fcnc.2} can be diagonalized
by a biunitary transformation for the left-handed and right-handed
components of fermion fields
\bea
f^\s{L}_i = \sum_k U^{f,\s{L}}_{ik} f^{'\s{L}}_k;\qquad
f^\s{R}_i = \sum_k U^{f,\s{R}}_{ik} f^{'\s{R}}_k,
\label{ff.1}
\eea
thus leading to the absence
of tree-level scalar bosons mediated FCNCs.

In what follows we will assume that the coupling of the scalar resonances with the fermions is proportional to the fermion mass, so that the generic coupling will be of the form ${\cal Y}m_\s{\rm f}$; however notice that the suppression of FCNCs is unrelated to this assumption.

Finally, the mass terms for the fermions are generated by using the 
bleaching counterparts of the mass eigenstates $f^{'\s{L}}$ and $f^{'\s{R}}$ and exploiting linearity of the bleaching procedure
\bea
\sum_i m_i \overline{\tilde f^{'\s{R}}_i} \tilde f^{'\s{L}}_i + {\rm h.c.} =
\sum_i m_i  \sum_{p,q} U^{f,\s{L}}_{ip}
          U^{\dagger f,\s{R}}_{iq}  \overline{{\tilde f}^\s{L}_p}  {\tilde f}^\s{R}_q  +{\rm  h.c.}
\label{ff.2}
\eea

\section{\label{pheno}Phenomenological implications}

\subsection{$X_0$ Decays into $W$ and $Z$ }

The tree-level NLSM widths for these decays read
\bea
\G(X_0\to WW)&=&\frac{g^2}{64\pi}\frac{g^2v^2}{4}\frac{\MX0^3}{\MW^4}\sqrt{1-x_\s{W}}\left(1-x_\s{W}+\frac34x_\s{W}^2 \right),\nonumber \\
\G(X_0\to ZZ)&=&\frac{g^2}{128\pi}\frac{g^2v^2}{4\cw^4}\frac{\MX0^3}{\MZ^4}\sqrt{1-x_\s{Z}}\left(1-x_\s{Z}+\frac34x_\s{Z}^2 \right),
\eea
where  $x_\s{V}=4M^2_\s{V}/\MX0^2$ with $V=W,\ Z$; the usual SM result can be recovered setting $\MW=\frac{1}{2}gv$ and using the Weinberg relation $\MW=\MZ\cw$. 

Since $X_0$ is to be identified with the ATLAS/CMS resonance, and therefore its mass is set to be equal to roughly $125$ GeV,  this decay is not energetically allowed, and processes in which one or  both gauge bosons are off-shell, \ie \mbox{$X_0\to VV^*\to V\bar f f$} and \mbox{$X_0\to V^*V^*\to \bar f f \bar f f$}, have to be considered. 

 However, 
in the NLSM there are four-fermion processes competing with these ones
and 
involving diagrams with off-shell scalar resonances, e.g. 
$X_0 \rightarrow X^{+'*} X^{-'*} \rightarrow 4f$, 
$X_0 \rightarrow X_3^{'*} X_3^{'*} \rightarrow 4f$,
$X_0 \rightarrow X_3^{'*} Z^* \rightarrow 4f$ and so on.
These diagrams have no SM counterpart even at leading order;
as a consequence, to obtain an estimate of the 
full $X_0$-width and of the different branching ratios,
a dedicated computation is needed 
in order to take into account the different NLSM background  w.r.t. the SM case. This lies beyond the scope of the present paper, and will be left for a later study.

Our present inability of evaluating these decays, leaves  us with the problem of fixing the SSB \vev~ parameter $v$; however, a suitable estimate of this quantity, valid at the approximation level appropriate for the ensuing analysis,  can be obtained by using the experimental value of the $Z$ mass and by assuming the SM tree-level relation
\be
\MZ=\frac{ev}{2\sw\cw}.
\ee
Then by replacing $e=\sqrt{4 \pi \alpha(0)}, \alpha(0) \sim 1/137$ 
and $\sw^2 \sim 0.23$,  we get
the value
\be
v=254\ {\rm GeV}.
\ee

Evidently, a similar analysis could be carried out for $\MW$; however
we choose to work on $\MZ$ since the diphoton decay channel we are going to analyze next has no dependence
on the Z-mass, the experimental value of which can therefore be used to get the needed estimate for $v$.

A comment is in order here. One might try to get additional information on the parameters of the nonlinearly realized electroweak theory by
exploiting the electroweak precisions data at LEP~\cite{ALEPH:2005ab}.
However, in the LEP experimental fit, the $Z$-$\gamma$ interference term
is evaluated by assuming that at tree-level the
$\rho$-parameter is equal to one~\cite{ALEPH:2005ab}. This in turn would introduce
a bias in the theoretical fit against the nonlinearly realized electroweak
model, by setting to one the parameter controlling 
the distinctive signature of the nonlinearly realized model, where
two mass parameters for the gauge bosons are allowed, unlike in the
SM case. 

Therefore it seems more appropriate to use the LHC data to
assess the effect of the second mass parameter (as e.g. in the parameterization proposed in~\cite{LHCHiggsCrossSectionWorkingGroup:2012nn} or \cite{Passarino:2012cb}), and then, as a second
step, to include also the LEP data in a more refined analysis.

\subsection{$X_0$ Decays into Photons}

\begin{figure}[!t]
{\white aaa}\includegraphics[scale=.775]{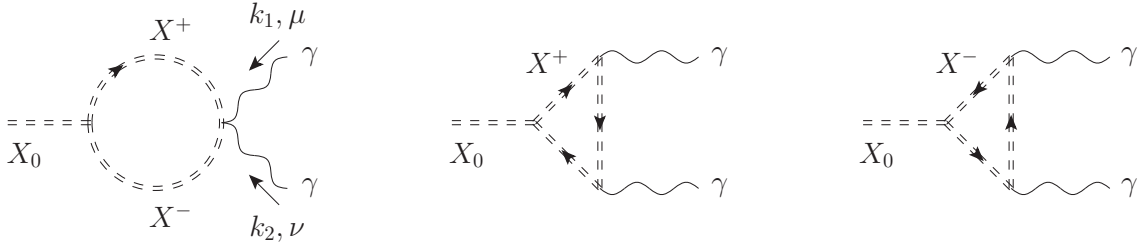}
\caption{\label{X0gg}The genuine NLSM diagrams contributing to the $X_0\to\gamma\gamma$ process. The remaining diagrams are equivalent to the usual SM graphs and  therefore are not shown.}
\end{figure}

A much cleaner channel for carrying out a first test of the NLSM
is the diphoton channel,  $X_0\to\gamma\gamma$, for which we show in \fig{X0gg} the purely NLSM one-loop diagrams contributing to this process.

After separating the NLSM new scalar contributions (${\cal A}_\s{\rm NL}$) from the fermionic (where we keep only the top contribution) and the vector-Goldstone-ghost ones (${\cal A}_\s{\rm f}$ and ${\cal A}_\s{\rm gauge}$ respectively), we get the following results

\bea
{\cal A}_\s{\rm NL} & = & - \frac{4e^2}{ \pi^2}
\frac{1}{M_{X_0}^2 \MW^2 s_W^2 v} 
\frac{1}{ e^2 v^2 - 4 \MW^2 s_W^2}
\Big [M_{X_0}^2
\Big (  \MW^2 s_W^2 - \frac{e^2}{4} v^2 \Big )^2 -
e^2 \MW^2 M_{X^\pm}^2 s_W^2 v^2 \Big ] \nonumber \\
& \times  &( k_1\cdot k_2\, g^{\mu\nu} -  k_1^\nu k_2^\mu )\epsilon^*_\mu(k_1)\epsilon^*_\nu(k_2)\left[ 1 + 2 M_{X^\pm}^2 
C_0(0,0,M_{X_0}^2,M_{X^\pm}^2, M_{X^\pm}^2,M_{X^\pm}^2) \right ],%\label{amp.nl}
\nonumber\\
{\cal A}_\s{\rm f} & = & \frac{e^3 }{\pi^2}\frac{ m_t^2 {\cal Y}}{M_{X_0}^2 \MW s_W}   ( k_1\cdot k_2\, g^{\mu\nu} -  k_1^\nu k_2^\mu )\epsilon^*_\mu(k_1)\epsilon^*_\nu(k_2)\nonumber \\
& \times & \left [-2 + (M_{X_0}^2 - 4 m_t^2)C_0(0,0,M_{X_0}^2,m_t^2,m_t^2,m_t^2) \right], 
%\label{amp.ferm} 
\nonumber \\
{\cal A}_\s{\rm gauge}  &=&  -\frac{e^4}{16 \pi^2} \frac{v}{M_{X_0}^2 \MW^2 s_W^2}
( k_1\cdot k_2\, g^{\mu\nu} -  k_1^\nu k_2^\mu )\epsilon^*_\mu(k_1)\epsilon^*_\nu(k_2) \nonumber \\
& \times & \left [ - M_{X_0}^2 - 6 \MW^2 + 6 (M_{X_0}^2 - 2 \MW^2) \MW^2 C_0(0,0,M_{X_0}^2,\MW^2,\MW^2,\MW^2) \right ].
\label{amp.gauge}
\eea  
In the formulas above $k_{1,2}$ are the momenta of the photons (with $2k_1\cdot k_2=\MX0^2$) and $\epsilon$ represents the photon polarization vector; notice that the appearance of the common tensorial structure  $ k_1\cdot k_2\, g^{\mu\nu} -  k_1^\nu k_2^\mu $ is dictated by gauge-invariance. Finally,
for the particular kinematic configuration of this decay the Passarino-Veltman three-point function $C_0$ is known to be (see, \eg\cite{Marciano:2011gm})
\bea
C_0(0,0,m^2,M^2,M^2,M^2) = -\frac{2}{m^2} f \left ( \frac{4 M^2}{m^2} \right)
\eea
where
\begin{equation}
f(\beta) =  \left  \{
\begin{tabular}{ll}
$\arcsin^2(\beta^{-\frac{1}{2}})$ & \text{for  $\beta \geq 1$}\\
$-\frac{1}{4} \Big [ \ln \frac{1 + \sqrt{1-\beta}}{1 - \sqrt{1-\beta}} - i \pi \Big ]^2$  & \text{for  $\beta < 1$}\\
\end{tabular} 
\right. 
\medskip
\end{equation}

It can be easily checked that setting $\MW=\frac{1}{2}gv$, $\MX0 = M_h$ (the Higgs mass) and ${\cal Y}=1$ the two amplitudes ${\cal A}_\s{\rm f}$ and ${\cal A}_\s{\rm gauge}$ reduce to their SM counterparts.
Also notice that ${\cal A}_\s{\rm NL}$ diverges for $\MW=\frac{1}{2}gv$, which is equivalent to $A \rightarrow 0$. This corresponds to the singularity associated 
to the change in the number of degrees of freedom, since for $A \rightarrow 0$
the $\chi$'s become the Goldstone fields and therefore one must not add the 
amplitude ${\cal A}_\s{\rm NL}$, since the Goldstone loop is already
included in ${\cal A}_\s{\rm gauge}$ [see \1eq{mass.1}]. 

Next, in order to compare with the SM case let us construct the ratio
\be
R=\frac{{\cal A}_\s{\rm NLSM}}{{\cal A}_\s{\rm SM-like}(a,c)},
\label{Rratio}
\ee
where ${\cal A}_\s{\rm NLSM}={\cal A}_\s{\rm NL}+{\cal A}_\s{\rm f}+{\cal A}_\s{\rm gauge}$, while ${\cal A}_\s{\rm SM-like}(a,c)$ is a linear combination of the
SM vector-Goldstone-ghost and fermionic contributions weighted, respectively, by two coefficients $a$ and $c$ which represent common rescaling factors with respect to the SM prediction for the Higgs couplings to vector bosons and fermions. The latter coefficients have been determined through fits to the LHC data \cite{Giardino:2012dp}.

 A comment is in order here. To carry out the fit to LHC data in a fully satisfactory way, one should make the comparison at the level of widths and cross sections. This is because, in addition to the dependence on the model parameters entering into the ratio $R$,  one should also consider 
an extra dependence arising 
from the $gg \rightarrow X_0$ cross section.
An estimate of such a dependence is left for a future study;
the ensuing preliminary discussion only aims at estimating the impact arising from the additional NLSM terms in the $X_0 \rightarrow \gamma \gamma$ amplitude.

According to Ref.~\cite{Giardino:2012dp},  one finds two possible scenarios that allow for an enhancement of the diphoton decay channel. The first scenario has the scalar coupling to fermions reduced with respect to the SM predictions and a somewhat enhanced Higgs boson couplings to vectors; the second scenario has the scalar coupling to fermions with opposite sign with respect to the SM prediction, as well as smaller couplings to the gauge bosons.    
The central values for these scenarios are roughly $(1.1,0.75)$ in the first case and $(0.8,-0.75)$ in the second case; clearly $(1,1)$ corresponds to the conventional SM case.

\begin{table}
\begin{tabular}{|c|c|c|c|c|c|c|}
\hline
$\alpha(\MX0)$ & $e$ & $\sw^2$ & $v$ & $\MX0$ & $\MW$ & $m_t$ \\
\hline\hline
$1/128$ & $\sqrt{4 \pi \alpha(\MX0)}$ & 0.23 & $254$ GeV & $125.6$ GeV & 80.385 GeV & 173.5 GeV \\
\hline
\end{tabular}
\caption{\label{inputpar}Input parameters for the calculation of the ratio $R$ of~\1eq{Rratio}. The choice \mbox{$\alpha(\MX0) \sim 1/128$} was taken from~\cite{Degrassi:2003rw}.}
\end{table}

The strategy we adopt is then the following. By fixing the parameters $(a,c)$ to any of the values quoted for the three scenarios, one automatically fixes the  normalization factor ${\cal A}_\s{\rm SM-like}(a,c)$. At this point we impose the condition $R=1$ and solve it in order to determine the corresponding value of the NLSM Yukawa coupling ${\cal Y}$. 

\begin{figure}[!b]
\includegraphics[scale=0.75]{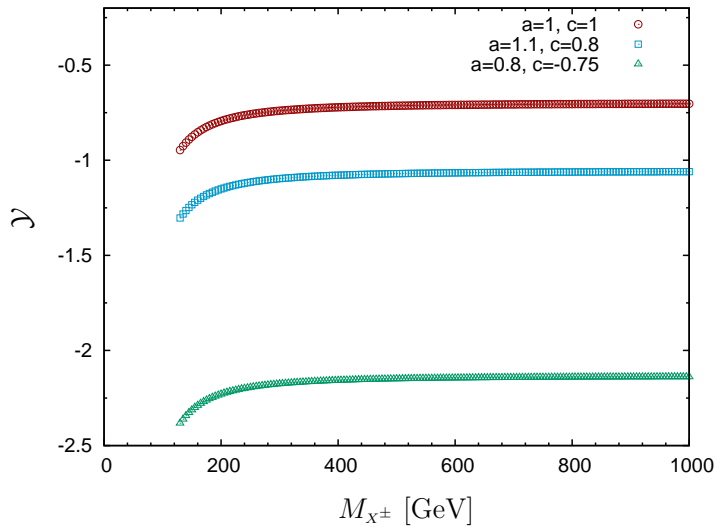}
\caption{\label{X0ggGraph}The NLSM Yukawa coupling needed for reproducing the scalar diphoton channel. The predicted Yukawa value is always negative.}
\end{figure}

The results of this procedure, taking as input parameters  the ones summarized  in Table~\ref{inputpar}, are shown in \fig{X0ggGraph}, where one can see that in any case the NLSM has the very distinctive signature of always requiring a negative Yukawa coupling between the top and the $X_0$ scalar to accommodate for the measured branching ratio in the diphoton channel.

\section{\label{concl}Conclusions}

The experimental verification of the Higgs mechanism is a two-step process. The first of these steps has been spectacularly completed by ATLAS and CMS with the discovery of a bosonic resonance at about $125$ GeV. The second step (and probably the most difficult one) crucially relies on the measurement of the new particle couplings and their comparison with the (very constrained) SM predictions for the Higgs boson.
We are therefore on the verge of a very delicate turning point, in which the experimental data should be checked not only against the SM (and, implicitly, all its underlying assumptions) but also against  all the  available theoretical scenarios that predict deviation from~it.

In this paper we have proposed a {\it plain vanilla} realization of one of these latter models, in the form of an electroweak SU(2)$\otimes$U(1) theory (dubbed NLSM) in which the gauge group is realized non-linearly, and the intermediate gauge bosons acquire their mass through the St\"uckelberg mechanism. 
Though non-renormalizable, the model is unitary and in addition the two simultaneous requirements of \n{i} satisfying the WPC (which singles out uniquely the field theory Hopf algebra) while \n{ii} being able to reproduce well-established experimental results (\eg the absence of FCNCs)  strongly constraint the form of the NLSM Lagrangian.

In this paper we have analyzed some peculiar features of the NLSM:

\begin{itemize}

\item When one tries to include scalar resonances, the minimum number of particles one can introduce in the NLSM is {\it four} corresponding to two neutral (one CP even and one CP odd) and two charged scalars: no scalar singlet is allowed. This fact {\it alone} singles out the NLSM  w.r.t. two of the most popular SM extensions, namely the two Higgs-doublet model and the Minimal Supersymmetric Standard Model, both requiring, in their minimal realization, {\it five} scalars;

\item To account for the (experimentally verified) suppression of the decay channel to VV  w.r.t  the $\gamma\gamma$, demands  the inclusion of a SSB mechanism {\it on top of} the St\"uckelberg mechanism. However, contrary to the conventional case, in the NLSM the \vev~ $v$ gives mass only to the CP-even resonance $X_0$ while all masses of the remaining particles and their respective couplings are unaffected, a fact which can account for  physics beyond the SM. 

\end{itemize}

Thus the NLSM constructed here represents, to the best of our knowledge, the first example of a model in which a SSB mechanism exists that has nothing to do with the generation of fermions and gauge bosons masses.

As a warming-up exercise towards a phenomenological analysis, we have also performed a preliminary study of the $X_0\to\gamma\gamma$ decay channel and found that,  in order 
to accommodate the ATLAS/CMS data, the NLSM CP-even scalar $X_0$ 
has always to couple to the top quark through a negative Yukawa coupling. 
And this is true  {\it even in the case in which the measured values would not ultimately deviate from the SM expected results}.

While it is definitely too early to envisage in the LHC measurements any clear hint of deviations from the SM predictions for the candidate SM Higgs, 
one might reasonably expect that in all scenarios the comparison with the NLSM could be
a very useful benchmark to pinpoint the Higgs mechanism
as the actual mass generation mechanism chosen by Nature.

\acknowledgments

We acknowledge useful discussions with A.~Vicini, K.~Ebrahimi-Fard and F.~Patras, and we thank S. Dittmaier for a critical reading of the manuscript.
One of us (A.Q.) is grateful to  the  ECT* for the warm hospitality. 
\appendix
\section{\label{app:a}Bleached Variables}

One can form local SU(2)-invariant variables (bleached fields)
as explained in~\cite{Bettinelli:2008qn}. The change of variables
from the original to the bleached fields is invertible.
Since the   bleached variables  are SU(2)-invariant, their hypercharge 
and electric charge coincide.

The SU(2) gauge fields $A_\mu = A_{a\mu} \frac{\tau_a}{2}$ and
the $\rm U(1)_\s{Y}$ gauge field $B_\mu$ are combined into the bleached
combination
\bea
w_\mu &=& w_{a\mu} \frac{\tau_a}{2} \nonumber \\
&=& \Omega^\dagger g A_\mu \Omega + g' \frac{\tau_3}{2} B_\mu 
+ i \Omega^\dagger \partial_\mu \Omega.
\label{a.g.1}
\eea 
One can easily verify that the above combination is invariant
under the $\rm SU(2)_\s{L}$-gauge transformations
\bea
\Omega ' &=& U_\s{L} \Omega , \\
A'_\mu &=& U_\s{L} A_\mu U_\s{L}^\dagger + \frac{i}{g} U_\s{L} \partial_\mu U_\s{L}^\dagger.
\label{a.g.2}
\eea
$w_{3\mu}$ represents the bleached counterpart of the $Z_\mu$ field, \ie
\bea
Z_\mu = \left . \frac{1}{\sqrt{g^2 + g'^2}} w_{3\mu} \right |_{\phi_a = 0} =
\cw A_{3\mu} + \sw B_\mu
\label{a.g.3}
\eea
where $\sw$ and $\cw$ are, respectively, the sine and cosine of the Weinberg angle, with
\bea
\sw = \frac{g'}{\sqrt{g^2 + g'^2}}; \qquad 
\cw= \frac{g}{\sqrt{g^2 + g'^2}},
\label{a.g.4}
\eea
and the photon $A_\mu$ is\footnote{Notice the change of sign in $\sw$ w.r.t.~\cite{Bettinelli:2008ey} in order
to match the conventions of~\cite{Denner:1991kt}.}
\bea
A_\mu = - \sw A_{3\mu} + \cw B_\mu.
\label{a.g.5}
\eea
The bleached counterparts of the $W^\pm$ fields are instead given by
\bea
w^\pm_\mu = \frac{1}{\sqrt{2}} ( w_{1\mu} \mp i w_{2\mu} ).
\label{a.g.6}
\eea

For a generic $\rm SU(2)$ fermion doublet $L = \pmatrix { u \cr d }$, bleaching yields 
\bea
\widetilde L = \Omega^\dagger L.
\label{a.g.7}
\eea

Finally, the bleached counterpart of the SU(2) scalar doublet $\chi = \chi_0 + i \chi_a \tau_a$ is given by
\bea
\widetilde \chi  = \Omega^\dagger \chi = \frac{1}{f} \left (\widetilde \chi_0 + i \widetilde \chi_a \tau_a \right),
\label{a.3}
\eea
with
\bea
\widetilde \chi_0 &=& \frac{1}{f} \left( \phi_0 \chi_0 + \phi_a \chi_a
\right ), \nonumber \\ 
\widetilde \chi_a &=&  \frac{1}{f} \left ( \phi_0 \chi_a - \chi_0 \phi_a + \epsilon_{abc} \phi_b \chi_c \right).
\label{a.4}
\eea
This results in two neutral scalar fields, one CP-even
($\widetilde \chi_0$) and one CP-odd ($\widetilde \chi_3$), and two
charged scalars $\widetilde \chi^\pm$, with
\bea
\widetilde \chi^\pm = \frac{1}{\sqrt{2}} \left(\widetilde \chi^1 \mp i \widetilde \chi^2\right).
\label{a.5}
\eea

\end{document}